\pdfoutput=1
  \documentclass[acmsmall,screen]{acmart}

\usepackage{booktabs} % For formal tables

\usepackage[ruled]{algorithm2e} % For algorithms

\SetAlFnt{\small}
\SetAlCapFnt{\small}
\SetAlCapNameFnt{\small}
\SetAlCapHSkip{0pt}
\IncMargin{-\parindent}

\usepackage{natbib}
\usepackage{soul}
\usepackage{hyperref}
\usepackage{cleveref}
\usepackage{balance}
\usepackage{mathtools}
\usepackage[utf8]{inputenc}
\usepackage[T1]{fontenc}
\usepackage{microtype}
\usepackage{caption, subcaption}

\def \class {\texttt{.class} }
\def \dex {\texttt{.dex} }

\setcopyright{rightsretained}
\acmPrice{}
\acmDOI{10.1145/3276494}
\acmYear{2018}
\copyrightyear{2018}
\acmJournal{PACMPL}
\acmVolume{2}
\acmNumber{OOPSLA}
\acmArticle{124}
\acmMonth{11}
\begin{document}
% Title portion. Note the short title for running heads
\title[ShareJIT: JIT Code Cache Sharing across Processes and Its Practical Implementation]{ShareJIT: JIT Code Cache Sharing across Processes and Its Practical Implementation} %in Android Runtime}

\author{Xiaoran Xu}
\affiliation{%
  \institution{Rice University}
%   \streetaddress{6100 Main St.}
%   \city{Houston}
%   \state{TX}
%   \postcode{77005}
   \country{USA}
}
\email{xiaoran.xu@rice.edu}
\author{Keith Cooper}
%\orcid{}
\affiliation{%
  \institution{Rice University}
%   \streetaddress{6100 Main St.}
%   \city{Houston}
%   \state{TX}
%   \postcode{77005}
   \country{USA}
}
\email{keith@rice.edu}

\author{Jacob Brock}
\affiliation{
	\institution{University of Rochester}
    \country{USA}
}
\email{jbrock@cs.rochester.edu}

\author{Yan Zhang}
\affiliation{
	\institution{Futurewei Technologies}
%     \streetaddress{2330 Central Expy}
%     \city{Santa Clara}
%     \postcode{95050}
    \country{USA}
}
\email{yan.zhang@huawei.com}

\author{Handong Ye}
\affiliation{
	\institution{Futurewei Technologies}
%     \streetaddress{2330 Central Expy}
%     \city{Santa Clara}
%     \postcode{95050}
    \country{USA}
}
\email{ye.handong@huawei.com}

\renewcommand{\shortauthors}{Xiaoran Xu, Keith Cooper, Jacob Brock, Yan Zhang, and Handong Ye}

\begin{abstract}
Just-in-time (JIT) compilation coupled with code caching are widely used to improve performance in dynamic programming language implementations. These code caches, along with the associated profiling data for the hot code, however, consume significant amounts of memory. Furthermore, they incur extra JIT compilation time for their creation. 
On Android, the current standard JIT compiler and its code caches are not shared among processes---that is, the runtime system maintains a private code cache, and its associated data, for each runtime process. 
However, applications running on the same platform tend to share multiple libraries in common. Sharing cached code across multiple applications and multiple processes can lead to a reduction in memory use. It can directly reduce compile time. It can also reduce the cumulative amount of time spent interpreting code. All three of these effects can improve actual runtime performance.

In this paper, we describe ShareJIT, a global code cache for JITs that can share code across multiple applications and multiple processes. We implemented ShareJIT in the context of the Android Runtime (ART), a widely used, state-of-the-art system.
To increase sharing, our implementation constrains the amount of context that the JIT compiler can use to optimize the code. This exposes a fundamental tradeoff: increased specialization to a single process' context decreases the extent to which the compiled code can be shared. In ShareJIT, we limit some optimization to increase shareability.
To evaluate the ShareJIT, we tested 8 popular Android apps in a total of 30 experiments. ShareJIT improved overall performance by 9\% on average, while decreasing memory consumption by 16\% on average and JIT compilation time by 37\% on average. 
\end{abstract}





\maketitle

\section{Introduction}\label{sec:introduction}
Runtime systems, execution engines and emulators for dynamic languages typically employ Just-in-Time (JIT) compilation on frequently executed sequences, and store the resulting compiled code in \textit{code caches} for the use of subsequent executions. This technique improves the execution speed of those hot sequences, but it also consumes significant amounts of memory and CPU resources to generate and store those code caches, along with the data structures to manage them. As JIT compilers have matured, they have been extended to include aggressive optimization, profiling and other features.
The resulting growth in code size and associated data structures, coupled with the presence of code caches in many distinct and simultaneous processes, has created a situation where the memory occupied by code caches and the CPU cycles used to manage them can degrade the system's overall performance. Since applications running on the same platform tend to share multiple common libraries, e.g. graphics libraries in Android, user-interface and animation libraries in JavaScript, and general software development frameworks, the opportunity for sharing cached code and amortizing costs exists.
However, the virtual machines (VM) that execute those applications are isolated, which forces repeated, independent compilation of shared libraries in multiple VMs, along with duplicate copies of the compiled code kept in process-private code caches.

This obvious drawback of the current software architecture motivates our work: sharing JIT code caches across applications. During the exploration of this idea, we have encountered several challenges. First of all, most JIT compilers leverage both runtime context and profile information to generate optimized code. The compiled code may be embedded with runtime-specific pointers, simplified through unique class-hierarchy analysis, or inlined recursively. 
Each of these "\textit{improvements}" can decrease the ``\textit{shareability}'' of JIT compiled code. Managing the trade-off between more highly optimized code and more shareable code is a serious problem. A second challenge for this work is choosing the granularity of code to share. 
If we choose to share at the class level, that will require changes to both the memory layout and management of the class-data area and the heap; those actions, in turn, will weaken the \textit{portability} of the sharing system.
If we choose to share at the method level, that will require a guarantee that the class data referenced by a method will be located consistently and correctly in each runtime system. 
Other challenges arise, including the need for new polices for updating the shared code cache and garbage-collecting it; determining if two runtime methods from different processes are functionally equivalent; and discovering the right timing to move from local interpretation to global sharing of compiled code.
This paper presents a new code caching and sharing architecture, ShareJIT, which addresses all of these problems mentioned. It describes one implementation, in the Android Operating System (OS), and several open implementation options.  The experiments section (Section~\ref{sec:experiments}) shows performance results from running ShareJIT on a collection of popular apps.

We decided to implement ShareJIT in Android due to both the ubiquity of Android devices and their resource constrained environments.  It appears that Android production apps are growing in size; at the same time, the market expansion of Android devices into developing countries is creating a base of systems with smaller RAM configurations.
These twin pressures make memory efficient execution of apps a pressing problem.  %Consumer Android devices currently have as little as 1-2 GB of RAM, and some devices in developing markets only have 512 MB.  
The problem is serious enough that Google is already targeting small-memory devices (as little as 512MB to 1GB of RAM) with a separate version of Android Oreo (Go Edition).
The Go Edition uses about half the memory footprint of Android N, and ships with slimmed-down versions of popular Google apps~\citep{androidgo,androidgo2}. 
% \citet{brock2018payjit} tested Android N on a Google Pixel phone and determined that it uses over 600 MB of RAM immediately after startup, and about 1.7 GB after 45 minutes of use. 
ShareJIT provides a partial answer to the problem of reducing memory pressure in app execution by identifying code that is used by multiple apps and allowing the apps to share JIT compiled versions of this code.  This approach has the added benefit of reducing JIT warm-up time in cases where an app can use a method that has already been compiled by another app.

\vspace*{.3cm}\noindent\textbf{Contributions: }
The main contributions of this paper are,
\begin{itemize} \setlength{\itemsep}{3pt}
\item A design of a global JIT code cache that shares across different processes, with minimal modifications to the VM layer memory layout. ShareJIT was designed to port easily to other versions of the runtime system. %\red{or saying ``easily composable/be assembled on any existing runtime system''?????} \blue{did I capture that sentiment? from xiaoran: yes!}
\item A set of policies to update and garbage-collect the global shared code cache, which coordinates and orchestrates all participating processes.
\item A cost model that analyzes the performance-critical parameters in sharing compiled code among inter-process methods.
\item A detailed implementation in Android 7 and evaluation on 8 widely used mobile applications that have billions of downloads. ShareJIT achieves an average of 16\% reduction in memory usage and an average of 9\% speedup in overall performance.
\end{itemize}

The rest of this paper is organized as follows. Section~\ref{sec:art} explains the background knowledge of Android Runtime system which is integral to the implementation of ShareJIT. Section~\ref{sec:implementation} describes the architecture and implementation of ShareJIT, including its key components, workflow, as well as garbage collection policies for the global shared cache. Section~\ref{sec:costmodelofsharing} builds a cost model for ShareJIT to analyze the performance-critical parameters, and examines how to set thresholds to most effectively share compiled code. Section~\ref{sec:experiments} demonstrates our experiments and discusses the results. Section~\ref{sec:related_work} outlines related work and Section~\ref{sec:conclusion} offers conclusions.

\section{The Android Runtime}\label{sec:art}
The Android Runtime (ART) system was introduced in Android 5 (Lollipop) as a new execution model for application code. Since then, the internal structure of ART has
evolved.
When we started the ShareJIT project, Android 7 (Nougat) was the newest version, so we used it to implement ShareJIT. In the rest of this paper, we use ``ART'' to refer to the runtime in \textit{Android 7.1.1\_r26}, a device-specific version of Android Nougat for the Google Pixel Phone. The following sub-sections briefly introduce background knowledge of ART to help demonstrate the implementation of ShareJIT in Section~\ref{sec:implementation}.

\subsection{Zygote Process}\label{sec:zygote}
As mentioned in Section~\ref{sec:introduction}, a large fraction of Android devices suffer from the tightening resource constraints. On such devices, launching an app can cause noticeable delays. To address this problem, ART provides a specialized process---\emph{zygote}, that spawns all other app processes. Zygote starts execution during the Android OS startup; it is initialized by preloading all the runtime Java classes and other shared resources that make up Android's rich frameworks. 
Because all other app processes start as a fork from zygote, they inherit zygote's memory and resources. We may think of zygote as a ``warmed-up'' process that speeds the startup of every other app's virtual machine. ShareJIT makes critical use of zygote's implicit resource sharing; it creates data structures for the global JIT cache in the zygote that are therefore shared to all other app processes. (See details in Section \ref{sec:internals}.)

\subsection{JIT Compiler and Cache} \label{sec:artjitcompiler}

\citet{brock2018payjit} provide a description of the JIT compilation policy and the code cache structure in Android N.  As background, we summarize that here.  First, ART compiles at method granularity instead of trace granularity (a trace is any series of instructions that may span multiple methods).  The advantage of this is a simpler compiler implementation, and reducing overhead for managing compiled code to achieve better performance~\cite{Inoue+:CGO11}.  The method-granularity compilation in ART makes sharing of compiled code feasible.  

There are three separate hotness thresholds used for JIT compilation.  After 5,000 method invocations or loop iterations, the method is ``warm'' and ART begins profiling the method and schedules it for ahead-of-time compilation when the device is idle and charging.  After 10,000, the method is ``hot'' queued for JIT compilation.  Finally, if a method's hotness reaches 20,000, it is likely to be in a hot loop.  Thus, the method is scheduled with high priority for on-stack replacement (OSR) compilation. The code produced by OSR is invoked by some specialized bytecode, which is typically a loop-closing branch. (The branch is replaced by code that invokes the OSR-compiled method.) Once the OSR compilation is completed, the interpreter will jump to the OSR-compiled code when it next crosses the specialized entry bytecode in the middle of the method. %\blue{did I mess up the explanation of how an OSR-compiled method is called?}\red{not at all}

ShareJIT does not share OSR-compiled code, because of the way that code is invoked.
Instead, ShareJIT limits sharing to code compiled at the ``hot'' threshold (10,000); that code presents a standard method-invocation interface. 

The JIT cache\footnote{When we refer to ``JIT cache'', we intend to both the JIT code cache and the JIT data cache, together.} initially allocates separate but adjacent 32 kB areas for code and data (stack maps and profile information for methods).
% exactly from PAYJIT - cannot use: The type \texttt{mspace} provides an interface for mallocing space, with limits on total size.
Whenever either space becomes full, garbage collection is triggered. Garbage collection occurs in two modes, partial collection and full collection, and they are performed alternately.  Partial collection removes non-entrant code and increases both code and data cache capacity equally.  Full collection additionally removes the profile information of methods which are warm but not yet hot, along with code that has not been executed since the last partial collection.  If the current JIT cache capacity is already at the maximum capacity (64 MB) and the collection could not free any space, the compiled code is discarded.

\subsection{From Dex Code to Machine Instructions}\label{sec:dexcode}
Android applications are often compiled from Java bytecode or directly from Java to ART's dedicated register-based \emph{dex} bytecode format~\citep{dexcode}. The dex file format~\cite{dexfile} (\dex) is structured differently from Java bytecode files (\class).
% Java bytecode, dex file format~\cite{dexfile} has different data structure \textcolor{red}{UNCLEAR} and organization with Java class file format.  
A \class file only contains one Java class. During runtime, the JVM will dynamically load the bytecode for each class from its corresponding \class file.  By contrast, a \dex file may contain all the classes of an Android app.  The constant pool is a per-class data structure existing in each Java \class file, but in a \dex file it is a global data structure shared by all classes. 
Thus, each symbolic reference in a \dex{} file is unique; if multiple classes reference the same constant, the \dex{} file will have a single copy of that value.
This is space-efficient but also has the side effect of limiting the number of references to $2^{16}$, since the global reference index is represented as an unsigned 16-bit integer. So from Android 5, ART started to support multiple \dex files for apps whose references are more than $2^{16}$~\cite{multidex}.

The global namespace of symbolic references in the dex file format leads to a different \textbf{symbolic reference resolution} procedure in ART compared with those in standard JVM.
Symbolic reference resolution is an optional phase of class linking in dynamic language execution.
It is a process for a dynamic language's virtual machine to locate classes, interfaces, fields, and methods referenced symbolically from a type's constant pool, and replace those symbolic references with direct references. Resolution only touches a type's constant pool entries but not the method bytecodes that reference them. But ART implements resolution as, (1) locating a global symbolic index referenced by a dex method when it's invoked; (2) storing the resolution result in a local, per-class data structure called the ``\emph{dex cache}'', which is similar to constant pool, but only contains resolved references; (3) replacing the global symbolic index used in the dex method bytecode with the local index in the dex cache. (Note that the runtime representation of class data in the JVM is immutable.)

Whether it is a global symbolic index or a local dex cache index, the ART JIT compiler will encode the index itself, but not the direct address it references, into the compiled code when generating machine instructions as long as there's no optimization associated to this index. (See details in section~\ref{sec:shareability})  When the compiled code is executed as a frame in the virtual machine stack, a reference to the runtime constant pool (dex cache) of the class of the current method is maintained, so that the real direct address can be located. Unresolved references will trigger a resolution request that will be handled by the runtime system. Under such code generation rules, ShareJIT only needs to guarantee that all the symbolic references that two dex methods use at runtime are the same when sharing compiled code from one method to another, while it does not care about the absolute direct addresses they reference in each virtual machine.

\section{The Design and Implementation of ShareJIT}\label{sec:implementation}

\subsection{ShareJIT Internals}\label{sec:internals}
\begin{figure}[t!]
	%\vspace*{-.5cm} % if we are not tight on space, do not pull this up vertically
	\includegraphics[keepaspectratio=true,scale=0.65]{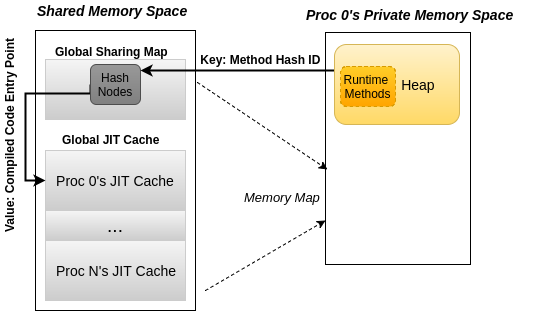}
 	\caption{The layout of the memory areas of ShareJIT}
\label{fig:internals}
\end{figure}

ShareJIT consists of two key components: a global JIT cache and a global sharing map. Figure~\ref{fig:internals} shows the layout of the memory areas of ShareJIT. The left side of the figure shows the shared memory space and the right side shows an example process's private memory space. Shared memories across processes are implemented as regions of Android Shared Memory (Ashmem), a component provided by Android OS to facilitate memory sharing. It is supported by the \textsc{POSIX} shared memory (Shmem) API in the kernel but wrapped with features to alleviate the low-memory pressure on Android devices. An Ashmem region is simply a memory segment backed by a file/device-driver, which can be mapped into the virtual address space of any process that has the file descriptor of that Ashmem region. 

ShareJIT uses the zygote process to create the shared memory space it needs during 
the zygote's startup (i.e. \emph{file\_descriptor = ashmem\_create\_region(name, size)}). 
Every app process inherits the resulting file descriptor when it is forked from zygote.
Then the shared memory is mapped into the app process' own virtual address space when that process' virtual machine is initialized (i.e. \emph{shared\_memory\_address = mmap (file\_descriptor, size, MAP\_SHARED,...)}\,). Note that since (1)~all the app processes
inherit the zygote's address-space layout, (2)~they all use the same startup sequence, and (3)~that process is deterministic, the shared memory area is mapped at the same virtual address by each participating virtual machine. This ensures that pointers to shared memory are valid across all processes. (Even in runtime systems which do not guarantee that the shared area is always mapped to the same virtual address across processes, ShareJIT can still locate objects inside the shared area through offsets between the base address of the shared area and the absolute address of those objects.)

\vspace*{.3cm}\noindent\textbf{The Global JIT Cache: }
The global JIT cache consists of consecutive, equal-size JIT cache segments owned by each app process.\footnote{The equal-size segment design simplifies parts of the ShareJIT implementation.
Because the shared cache is file-backed, demand-paged, and mapped into each app's virtual address space, the cost of enlarging the per-app segment size is minimal; it only incurs a physical memory cost if the space is used. Thus, our approach is to choose a "large enough" per-app segment size and avoid the
complications and costs that would come with managing different per-app segment sizes.}
The management and access-control of each segment is as follows.
\begin{itemize}
\item The owner process of a JIT cache segment has access to store compiled code in the segment.  Memory allocation for the code inside the segment obeys the original mechanism in ART.

\item Once compiled code is written into the cache segment, neither the owner process nor any other process may modify the code. 

\item The owner process may remove some compiled code during garbage collection, but only when it is not in use by any process. Memory deallocation for the code inside this segment also obeys the original mechanism in ART.

\item Non-owner processes only have the right to read and execute the code inside the segment.
\end{itemize}

By assigning each cache segment an owner and restricting other processes' access rights, ShareJIT carefully avoids the potential of read-write and write-write conflicts within the code cache; it also reduces the risk of memory attacks from security exploits that have low-level access to 
memory.\footnote{The problem of an attacker spoofing a code-cache entry is difficult; the best solution will depend on assumptions about the rest of the system. If the JIT is corrupted, then neither code cache policies nor implementation will prevent the attacker from taking over the system. Thus, we assume that the JIT is uncorrupted and that only the JIT has write access to the code cache. In this scenario, a hash key computed over the method bytecode should be sufficient to ensure a method's validity. (The hash key will be described in the next subsection.) 

If arbitrary code in the process can write into the code cache, then additional measures and 
additional computation would be necessary to detect a corrupted native-code segment. For example, 
the JIT could compute a secondary key over the native code and securely retain it in a 
map from hash key to secondary key. The 
JIT could, on demand from the prospective sharer, recompute the seconday key and compare it to
the value stored in its secure map. Such a scheme could provide an added degree of detection and
reassurance to the prospective sharer.}
%an write into the code cache, but less restrictive situations can be managed at the cost of computing additional information, such as summaries of the side effects of the code segment and a hash key computed over the JIT-compiled code. E.g., a summary of locations read, locations written, and total binary codes; a hash of that information should provide an added degree of reassurance. }
By reserving the memory allocation and deallocation mechanism of the original runtime system inside each segment, ShareJIT minimizes the modification to the VM-layer memory management, leading to strong portability.

Theoretically, ShareJIT is able to declare as many cache segments in the global JIT cache as necessary, because the memory mapping facility \texttt{mmap} supports demand paging.  This means the operating system copies a disk page into physical memory only if an attempt is made to access it and that page is not already in memory. Thus, the use of physical memory is determined by demand, rather than by the size of the backing file. But to be memory efficient, we set the maximum number of cache segments as $64$ in our implementation. After the number of app processes reaches this maximum count, ShareJIT offers two options for a newly started process: (1) reclaim a cache segment if its owner is dead (killed by the OS, e.g., Android low-memory killer~\cite{lowmemkiller}, or killed by the user); (2) if no pre-occupied cache segment is freed, create traditional JIT cache in this process's own private memory space, which does not communicate with the global JIT cache. Note that we implemented a ``lazy'' reclamation of a dead process's cache segment; another option is to let a process broadcast its death signal and be reclaimed by ShareJIT in an ``eager'' style.

\vspace*{.3cm}\noindent\textbf{The Global Sharing Map: } The global sharing map is the bridge from runtime methods in each virtual machine's heap to the compiled code in the global JIT cache. Whenever a method $m$ is hot enough and JIT compiled in some process $n$, $n$ is responsible for creating a hash node for $m$ in the global sharing map.
The key of the node is the \textit{hash-identification} of $m$ and the value is the entry point of the compiled code of $m$ in process $n$'s cache segment. We define the \textbf{hash-identification} of a method $m$ as the hash code of $m$'s method signature and byte-by-byte dex 
code.\footnote{We use a 128-bit hash code value in our implementation. According to our experiments, large apps such as Facebook have up to $10^6$ methods in total, while small apps tend to have around $10^4$ methods. Even if we have \mbox{1,000} the largest apps and $10^9$ methods in one system, the hash collision possibility is as low as $10^{-21}$, which is negligible.}
Section~\ref{sec:dexcode} proves the legitimacy of identifying functionally equivalent methods with such a hash-id, whether the symbolic references used in the dex methods are statically linked or will be dynamically linked.  A hash node also contains an unsigned integer reference count of the processes that are using the compiled code associated with this node. The reference count tells the garbage collector when the method's compiled code can be collected and freed. (See details in Section~\ref{sec:gc})

The global JIT cache is divided into contiguous segments, each with a single writer, to simplify both management and synchronization. This scheme does not work for the global sharing map. Each process that uses the global sharing map needs the right to read and write it. Since Ashmem does not have a serialization mechanism for a multitude of readers and writers in a concurrent system, we use a \textit{semaphore} to protect the global sharing map. 

\subsection{ShareJIT Workflow}\label{sec:workflow}
\begin{figure}[t!]
	%\vspace*{-.5cm}
	\includegraphics[keepaspectratio=true,scale=0.42]{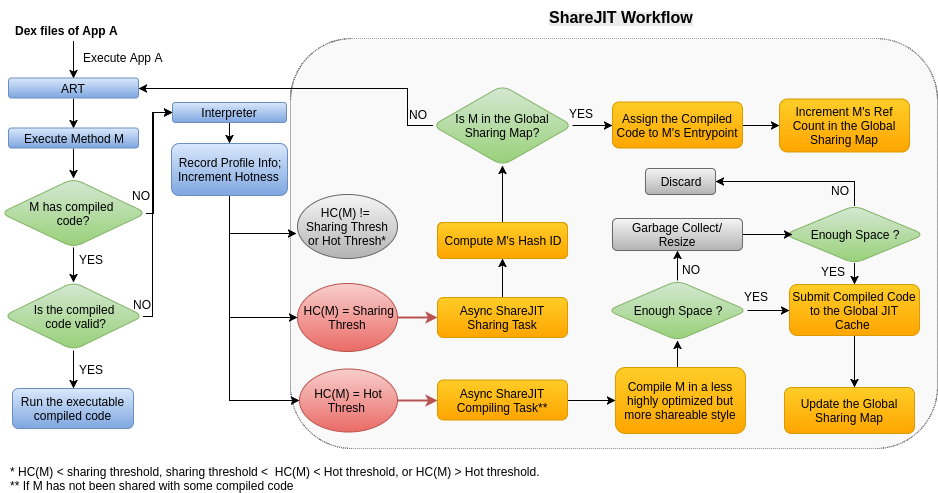}
 	\caption{The Work-flow of ShareJIT}
\label{fig:workflow}
\end{figure}

Figure~\ref{fig:workflow} shows the main workflow of ShareJIT. An Android app's lifetime starts with being installed, being launched which will trigger zygote to fork a child process for it and initialize a runtime system inside the process, and eventually being executed by the runtime system. When a method $M$ from the app $A$ is invoked, ART will first check if $M$ has corresponding compiled code as its execution entry-point. A method's entry-point could point to the JIT code cache area, the interpreter or null. (It could also point to the native code area, but we omit the scenario of native code execution in our discussion since it's irrelevant to the design and implementation of ShareJIT.) If $M$ has compiled code, ART will execute it directly from its entry-point; if not, ART will interpret it. After the interpretation, M's hotness count will be incremented; and if M is already warm, which means a profile was created and associated with it, the profile will also be updated according to the execution. 

Two thresholds govern the decisions in the ShareJIT workflow: the \textit{sharing threshold} and the \textit{hot threshold}. The \textbf{sharing threshold} determines how hot a method must be before ShareJIT will attempt to share it. The \textbf{hot threshold} determines how hot a method must be before ShareJIT will attempt to compile it. %\red{actually, hot threshold(10000) is the threshold to JIT compile a method, OSR threshold is 20000, we didn't discuss that here.}

These two thresholds divide the hotness count of some method $m$, HC(m), into five distinct ranges.
\begin{enumerate} \setlength{\itemsep}{3pt}
\item $HC(m) < sharing\ threshold$~: A value of $HC$ in this range triggers no immediate action.

\item $HC(m) = sharing\ threshold$~: When $m$ reaches the sharing threshold, ShareJIT will create a sharing task for it and add that task to the JIT thread's task queue. When the JIT thread reaches this task, it will (a) compute the hash-identification for $m$, (b) check if the global sharing map has code for this hash-identification, (c) update $m$'s code to use the compiled code if found, as well as incrementing the reference count of the corresponding hash node.

\item $sharing\ threshold < HC(m) < hot\ threshold$~: A value of $HC$ in this range triggers no immediate action.
\item $HC(m) = hot\ threshold$~: When $m$ reaches the hot threshold, if $m$ has not been shared with corresponding compiled code,  ShareJIT (a) compiles the code for $m$ in a shareable manner, (b) inserts it into the shared cache, (c) updates the local code in $m$ to use the newly compiled code, and (d) adds a new hash node for $m$ in the sharing map. 

\item$HC(m) > hot\ threshold$~: By the time a method reaches this range with its hotness count, it should have been shared with compiled code or compiled into the shared cache. If $m$ reaches the OSR threshold (20,000), it triggers an OSR compilation, but the resulting code is not shared by ShareJIT because of its non-standard entry provisions. We do not discuss the details of OSR mechanism since it is not in our scope.
\end{enumerate}
\noindent
ShareJIT uses the same hot threshold as ART's default JIT---that is, 10,000.
Sections~\ref{sec:costmodelofsharing} and~\ref{sec:sharingthreshold} explain how we set the sharing threshold.
Section~\ref{sec:shareability} describes some of the constraints placed on compilation of globally shared code mentioned in range (4).

\subsection{The Shareability of JIT Compiled Code}\label{sec:shareability}
JIT compilers often generate more highly optimized code than ahead-of-time (AOT) compilers do because JIT compilers have extra runtime context. But overuse of such runtime context can make the code unsharable across processes. For example, replacing symbolic references in method invocation instructions (i.e. \texttt{invoke-direct} instruction in dex bytecode) with the absolute addresses of callee methods can reduce function call overhead. (This transformation happens as part of method devirtualization driven by class-hierarchy analysis.) 
Unfortunately, embedding absolute addresses makes the code unsharable because the method bytecode is located in the class-data area and is likely to be at different locations in different processes.
Thus, ShareJIT disables such usage of process-specific runtime context. The alternative would be to require that ShareJIT translates a private pointer in one VM to the corresponding address in another VM, which is impractical.  

Another optimization that poses problems for ShareJIT is inline substitution. As discussed in Section~\ref{sec:dexcode}, ShareJIT only needs to ensure the symbolic references two dex methods use are literally identical to determine that sharing compiled code between them is legal, while it can ignore the content that those symbolic methods reference.
%from one to the other, without caring about the content they reference to. 
However, inlining a callee method means dereferencing the pointer in an invocation instruction, and replacing it with the actual method 
code.\footnote{In the ART JIT compiler, it inlines the intermediate representation (control flow graph) and then translates the two methods together.} 
So to share code that has been inlined, ShareJIT needs to ensure that all of the code in the inlined section is identical in both ``sharer'' process and ``sharee'' process.
Specifically, ShareJIT could record a list of the inlined callee methods, and check that each pair of callee method in the ``sharer'' process and the ``sharee'' process are also functionally equivalent. This mechanism would require extra space and time to compute and store the hash-identifications of callee methods; it would significantly increase the overhead of sharing, particularly when inlining is performed recursively, i.e. ART JIT compiler allows an inlined callee method to inline its own callee, up to 3-level depth.  

Because of this overhead, the current ShareJIT implementation only inlines built-in methods---methods that are defined in the classes preloaded by the zygote process during its initialization. It does not inline app-specific methods---methods defined in app's \dex files. This restriction guarantees that if a callee is inlined, its definitions in both the ``sharer'' and the ``sharee'' runtime system are identical; they are the same implementation inherited from their common super-process: the zygote.

ShareJIT sacrifices some potential performance gain from inlining. It could inline more callee methods at the cost of a decrease in the shareability of compiled code. We discuss how this reduction in inlining affects the overall performance of ShareJIT in Section~\ref{sec:results}.

\subsection{Garbage Collection} \label{sec:gc}
As described in Section~\ref{sec:artjitcompiler}, garbage collection occurs in two modes: partial collection removes non-entrant code and increases both code and data cache capacity equally; full collection additionally removes the profile information of methods which are warm but not yet hot, along with code that has not been executed since the last partial collection. Non-entrant code are the compiled code that has been discarded for various reasons, the most frequent of which is deoptimization from aggressively compiled code, e.g. bounds check elimination, dynamic type assertion, inline caching, etc.  JIT compilers could try these optimizations not only because they could gather more profile information of the input method, but also because they could afford possible deoptimizations if such too-aggressive optimization attempts end up failing during the subsequent method executions--it just switches back to the interpreter and let the garbage collector free the discarded code.

Unlike those optimizations described in Section~\ref{sec:shareability} that have to be disabled or restricted in ShareJIT for shareability, these aggressive optimizations could be reserved in ShareJIT with proper modification of the garbage collector. We will use inline caching as an example to explain how these optimizations are compiled, when the deoptimization would be triggered, and how the garbage collector should deal with the concomitant discarded compiled code in a global sharing JIT cache.

Inline caching~\cite{deutsch1984efficient} is an optimization technique to speed up runtime method binding. It remembers the results of previous virtual method lookups ``inline'', i.e. directly at the call site. Those results are called ``inline caches''. The concept relies on the empirical observation that the objects that occur at a particular call site are often of the same type. ART collects inline caches for every virtual callee method invocation in a caller method, and stores them in the JIT data cache as ``profile information'' for this caller method. The JIT compiler will perform this optimization when an inline cache is monomorphic or polymorphic~\cite{holzle1991optimizing} during compilation, and also emit a backup ``slow-path'' which leads to a normal unoptimized/safe execution flow. If the ``slow-path'' is executed, which means the type assertion optimization fails, a concomitant deoptimization exception will be thrown to the runtime system afterwards. 

The handler for deoptimization exceptions invalidates and discards the offending compiled code. It resets the hotness count of that method to zero. If the method becomes hot again in the future, a new version of compiled code will be generated. 
%we omit the discussion of the method will become heated before becoming hot, and we have a shared method list to prevent shairng a method a same version of compiled code that had been shared and deoptimized previously.
If this situation occurs soon enough, i.e., before the next collection can free the older version, (this situation arises often in practice,) the original JIT would simply use the newly compiled version and let the garbage collector free the old version because the old version is discarded by itself; however, ShareJIT must decide which version to keep in the global sharing map from a global perspective. 

To deal with this situation, we introduce the first rule for the ShareJIT garbage collector: \textit{``Newer is better.''} Because the newer compiled code results from richer and more precise execution information, ShareJIT always overwrites the old version in the global sharing map with the new one. To prevent ``sharee'' processes from executing the discarded old code, a validity check is put before any compiled code is executed as shown in Figure~\ref{fig:workflow}.\footnote{Another scenario where a process might attempt to access invalid compiled code is when the process which produced the compiled code died and its cache segment was reclaimed.} Another way to cope with discarded compiled code is to broadcast its invalidation, which allows the ``sharee'' processes to proactively delete that entry point. In either approach, the garbage collector can safely collect and free the old code.

We use inline caching as an example to introduce why and how the ShareJIT garbage collector should behave differently, not only because it's one of the common deoptimizations, but also because inline caches are part of the method profile information that is stored in the JIT data cache. (The runtime system starts to collect profile information for a method when it becomes ``warm''.) From the description of inline caches above, we could see that once a sharing relationship is built, the ``sharee'' method's profile becomes useless since it was intended to help in the JIT compilation but the ``sharee'' method will simply not be 
compiled.\footnote{If a ``sharee'' method deoptimizes from running the compiled code and subsequently becomes warm again, which happens infrequently, the method can re-collect the profile information while it is getting hot.} So the garbage collector could free the profiles of all ``sharee'' methods during a full collection. Thus, a ``sharee'' method has no associated memory in the JIT cache at all---compiled code and stack maps are never created, while the profile is deleted.  (Note that the JIT data cache consists of two parts---method profiles and stack maps, while JIT code cache only contains compiled code, so ShareJIT saves more data cache than code cache, which will be demonstrated in Section~\ref{sec:results}.)

The second rule for the collector is \textit{``Only collect compiled code that is not being used by any process''}. Recall that each hash node in the global sharing map has a reference count which records the number of ``sharee'' processes for this node. The garbage collector will only collect and free the compiled code that has a reference count of~0. This policy might reduce the amount of code ShareJIT would collect compared with the default JIT, but it's beneficial to the performance. 

Detailed pseudo code is shown in Algorithm~\ref{alg:gc}. The variable $own\_method\_map$ holds the methods whose compiled code are stored in a process's own cache segment, while $sharee\_method\_map$ holds the methods whose compiled code is shared from other processes' cache segments. Traverse the $own\_method\_map$, if some compiled code $c$ is not its corresponding method $m$'s entry point, and the reference count of $m$ in the global sharing map is 0, delete it in the $own\_method\_map$, and also delete it in the global sharing map; traverse the $sharee\_method\_map$, if some compiled code $c$ is not its corresponding method $m$'s entry point, delete it in the $sharee\_method\_map$ and decrement the reference count of $m$ in the global sharing map. Note that for a method $m$ in the $sharee\_method\_map$, even if its reference count $RC(m)$ becomes zero after being decremented, the collector is not allowed to delete the node in the global sharing map or to free the compiled code $c$, because ShareJIT only grants the owner process the privilege to allocate/deallocate memories in its cache segment.

\begin{algorithm}
\vspace{.15cm}
\For{every pair (compiled code c, method m) in own\_method\_map }{
	\If{c is not m's entry point and RC(m) == 0}{
    	delete (c, m) in own\_method\_map\;
        delete m in global\_sharing\_map\;
        free c;
    }
}
\For{every pair (compiled code c, method m) in sharee\_method\_map }{
	\If{c is not m's entry point}{
    	delete (c, m) in sharee\_method\_map\;
        decrement RC(m) in global\_sharing\_map\;
    }
}
\caption{Checking Method Entry Points in a Partial or Full Garbage Collection}
\label{alg:gc}
\vspace{.15cm}
\end{algorithm}

\section{Understanding the Costs of {ShareJIT} Enabled Execution} \label{sec:costmodelofsharing}
ShareJIT introduces complex behavior that affects the runtime performance of apps.
To understand that behavior, it helps to consider three different perspectives on how the shared global JIT cache affects performance and memory utilization, as described in the following three subsections. In each case, the performance improvement is difficult to predict or model because it depends in a detailed way on both the future behavior of each process and on the overlap between processes.

\subsection{Single-Process Performance}  
From the perspective of a single executing app, ShareJIT has three direct effects.
\begin{enumerate} \setlength{\itemsep}{0pt} \setlength{\parskip}{3pt}
\item Once a method $m$ has been invoked \textit{sharing threshold} (ST) times, the process executing the app checks to see if $m$ has corresponding compiled code in the global shared cache. Specifically, the running process computes a 
global hash-identification for $m$, and checks in the global sharing map for an instance of
that hash-identification. 
\item If the running process discovers a shared, compiled copy of $m$, it links the runtime system's call to the shared, compiled implementation, and updates the reference count of $m$ in the global sharing map.
\item If the lookup fails to find a shared, compiled copy of $m$, the running process continues to interpret $m$ until $HC(m)$ reaches the \textit{hot threshold} (HT).
At that point, the executing process schedules a JIT compilation of $m$.
\end{enumerate}

\noindent Each of these actions has a cost. From the single-process perspective, the search for a shared, compiled copy of $m$ is an added cost.
But if the search finds a copy of $m$ in the shared global cache, the cost pays off, because the running process begins to execute compiled code earlier than it would with the standard Android JIT, and additionally avoids the cost of JIT-compiling $m$. 

Defining $\Delta{}T(m)$ as the reduction
in execution time from running the compiled version of $m$ rather than the interpreted version,
then it should save at most $\Delta T(m) \textit{(HT - ST)}$ time. Because the standard JIT would have interpreted $m$ until $HC(m) \geq HT$ while ShareJIT started executing compiled
code when $HC(m) \geq ST$. 
Once $HC(m)$ exceeds $HT$, the runtime cost of executing $m$ \textit{approximates}
the cost that would occur with the standard Android JIT. We say it approximates the cost 
because ShareJIT restricts optimizations in the shared compiled code in a way that the standard JIT does not.

Thus, the factors in ShareJIT execution that may slow the app are limited: 
\begin{enumerate}\setlength{\itemsep}{0pt} \setlength{\parskip}{3pt}
\item the cost of computing the global hash-identification for a method $m$, defined as $H(m)$,
\item the cost of checking the global sharing map for $m$, defined as $L(m)$,
\item the cost of linking the call site to the shared, compiled copy of $m$,
\item the cost of updating the reference count for $m$ in the global sharing map,
\item and any lost efficiency from restricted optimization.
\end{enumerate}
Of these costs, the first two are the most significant because they are incurred every time some
method crosses the sharing threshold and does a lookup in the global sharing map. While the cost is trivial for a single method, in aggregate, it occurs many times.
Items (3) and (4) only occur if that lookup succeeds: a much smaller number.
The lost efficiency from restricted optimizations is a default cost---the sum of all JIT compiled methods' performance degrading, which is not correlated with the sharing attempt. 

Note that in the implementation of ShareJIT in Android, the cost of computing method hash-identifications is paid at runtime. For other virtual machines in which the runtime representation of class data including bytecode is immutable, such as a standard JVM, method hash-identifications could be generated at compile time and stored in \class files. One drawback of this approach is that it consumes extra runtime space after these \class files being loaded. 
ShareJIT allows an implementation flexibility as to trading-off between time cost and space cost at runtime for identifying methods globally. In ART, computing method hash-identifications at runtime is the only choice as explained in Section \ref{sec:dexcode}. Besides, Android devices are already suffering from an increasing memory pressure.

On the other hand, ShareJIT has two major sources of improvement: avoiding the cost of JIT-compiling the method $m$ and the savings from additional compiled executions up to $(HT - ST)$ times.
As might be expected, the total number of executions of $m$ in the app determine the expected 
improvement. Assume that $m$ already exists in the shared code cache. 
If $HC(m)$ stops at $ (ST + 1)$, then the impact will be negligible.
As $HC(m)$ grows from \textit{ST} to \textit{HT}, the savings from 
executing compiled code will increase.
Once $HC(m)$ exceeds \textit{HT}, the savings from 
executing compiled code stop, and the only additional improvement comes from the fact that ShareJIT avoided the cost of JIT-compiling $m$.

\subsection{System-Wide Performance}
From the perspective of the overall system, ShareJIT will fundamentally change the relationship
between the amount of time spent JIT-compiling code and the number of method invocations.
With ShareJIT, a single compile step can improve the performance of multiple processes and multiple 
apps. With ShareJIT, a single app can begin executing compiled code at an earlier point in its
progress than it would with the standard Android JIT---earlier by 
(\textit{HT} - \textit{ST}) method invocations.

The impact of this effect depends, heavily, on how invalidations in the global cache are handled.
The benefit from cross-processes sharing depends on the compiled code remaining in the shared cache.
As explained in Section~\ref{sec:internals}, the current implementation keeps a process's image in the shared cache intact until it must be reclaimed. That is, if process $p$ compiled and cached method $m$ and, subsequently, exited, $p$'s copy of the compiled code for $m$ will remain in the shared cache until some newly spawned process needs a segment \textit{and} $p$'s segment is the next segment to be reclaimed. This policy maximizes the residency of shared code in the cache and increases the impact of the shared cache on overall performance. Although recompilation resulted from deoptimization also causes shared cache invalidation and update, it happens infrequently on shared code in practice, and sharing relationship can be rebuilt again with the recompiled code quickly.

To summarize the analysis above, %if a method $m$'s hotness $HC(m)$ stops growing at a point below the sharing threshold, then the cost of executing $m$ is the same in ShareJIT and default JIT. For other methods,
we could build a cost model for executing a method $m$ as: (in terms of the difference between ShareJIT and default JIT)
\begin{equation}
\label{eq:benefit}
F(m) = 
\begin{cases}
0, & \text{if } HC(m) < ST \\
-H(m) - L(m) + S(m)\Delta T(m)(HC(m) - ST), & \text{if }\ ST \leq HC(m) < HT \\
-H(m) - L(m) +S(m)\Delta T(m)(HT - ST) + S(m)J(m) , & \text{if } HC(m) \geq HT 
    \end{cases}  
\end{equation}

\noindent where $H(m)$, $L(m)$, $\Delta T(m), HC(m),\ ST$ and $HT$ have been defined before; $S(m)$ returns a binary value 1 or 0 depending on whether the lookup in the shared global cache for $m$ succeeds or not; $J(m)$ is the JIT compilation cost for $m$. (To simplify, we ignore the difference in $J(m)$ between ShareJIT and default JIT.) Then $F(m)$ would be the ultimate performance improvement for $m$ we could gain from ShareJIT, compared with default JIT.

$\Delta T(m)$, the speedup from executing the compiled code of $m$ versus interpreting $m$, is computable at runtime, although we didn't implement the calculation. Given $Ti(m)$, time cost of interpreting $m$, and $Tc(m)$, time cost of running the current version of compiled code of m, then obviously $\Delta T(m) = Ti(m) - Tc(m)$. $Ti(m)$ and $Tc(m)$ could be recorded during the execution of $m$.

However, the final hotness of a method, $HC(m)$, is unknowable. We cannot precisely predict how many times a method $m$ will be invoked in the future. We cannot predict if the compiled code will become invalid due to the recompilation of $m$ or the reclamation of the ``sharer'' cache segment, either. So we are not able to analyze the performance gain from Equation~\ref{eq:benefit} through mathematical modeling. Then when is the best time to share?

The similar problem motivated all the studies for JIT compilation policy in academic history---``when is the best time to JIT compile a hot method?'' The mainstream of JIT compilation policy in industry is using a hotness threshold or a method invocation threshold, e.g. ART, HotSpot JVM server compiler~\cite{Paleczny+:JVM01}, and IBM JIT compiler~\cite{suganuma2000overview};
methods whose hotness counts or invocation counts reach the threshold will get compiled at runtime. 
It is both easy-to-implement and efficient. 
Applying that heuristic on sharing policy, we assumed a single sharing threshold $ST$ of method hotness, at which ShareJIT attempts to share, in the workflow of ShareJIT and the analysis above. The total system performance improvement ShareJIT gains compared with default JIT are:
      
\begin{equation}
\label{eq:total_benefit}
Y(ST) = \sum_{ \mathclap{HC(m) \geq ST}} F(m) =
-\sum_{ \mathclap{HC(m) \geq ST}} C(m) +  \sum_{ \mathclap{HC(m) \geq ST}} G(m)
\end{equation}

\noindent where $C(m)$ is simplified for $H(m) + L(m)$, and $G(m)$ is simplified for the positive terms in Equation~\ref{eq:benefit}, despite of the value of $HC(m)$. We can see that the major tradeoff comes from the \textbf{warming methods}, those with $ST \leq HC(m) < HT$. For these warming methods, the benefit is the sum of inevitable increased cost from checking more methods and potential increased gains from sharing methods earlier and sharing more methods, as $ST$ decreases.
The key to maximizing total benefit is to find the best sharing threshold $ST$ for this term, given
as $Y(ST)$ above.  Section~\ref{sec:sharingthreshold} describes our experiment to find this value.

\subsection{System-Wide Memory Utilization} \label{sec:memory_utilization}
From the perspective of system-wide memory utilization,
the effect of ShareJIT is conceptually simple. ShareJIT reduces system-wide memory use for compiled-code
in proportion to the amount of sharing and the size of those methods. In practice, we can measure total
memory use for compiled code with the standard Android JIT and with ShareJIT and subtract. 
To make predictions, however, is difficult because the actual savings depend on how different apps 
overlap in time and how the eviction policy for the shared cache works.

In practice, we measured this effect and found that ShareJIT
reduced overall JIT cache use by roughly~16\%.  (see Section~\ref{sec:experiments} for more discussion)

\section{Experiments}\label{sec:experiments}
\subsection{Benchmarks}
To evaluate the performance of ShareJIT compared with the default JIT, we chose 8 widely used apps as benchmarks: \textit{Airbnb, Amazon, Chrome, Facebook, Firefox, Instagram, Googlemaps, Skype}. The specific version and release date information are listed in table \ref{tbl:apps}.
\begin{table}[b]
\centering\tabcolsep=9pt\renewcommand{\arraystretch}{1.25}
\caption{Apps Used in Experiments}
\label{tbl:apps}
\begin{tabular}{|l|l|l|}
\hline
\textbf{APK} & \textbf{Version} & \textbf{Release Date}\\\hline
\it{}Airbnb       & 17.50          & Dec. 16, 2017  \\ \hline
\it{}Amazon       & 12.7.0.100       & Aug. 28, 2017   \\ \hline
\it{}Chrome       & 57.0.2987.97      & March 9, 2017   \\ \hline
\it{}Facebook     & 100.0.0.20.70     & Oct. 18, 2016  \\ \hline
\it{}Firefox       & 52.0-2015474475       & March 3, 2017   \\ \hline
\it{}Instagram    & 10.25.0-60813718       & June 9, 2017   \\ \hline
\it{}Googlemaps    & 9.61.1-961102122   & Sep. 12, 2017   \\ \hline
\it{}Skype        & 8.12.0.14        & Dec 11, 2017   \\ \hline
\end{tabular}
\end{table}
They were chosen for three primary reasons:
\begin{itemize} \setlength{\itemsep}{3pt}
\item They are all mass-market apps in real life; (Except for \textit{Firefox}. \textit{Firefox} was chosen because it's in the same category with \textit{Chrome}--they are both web browsers, and we would like to see if apps in the same category tend to share more code in common.)
\item They all have easy-to-use graphical interfaces which can be automated by our scripts to imitate the real users' activities;
\item They all have reasonable number of JIT events during run time. (Games are also heavily downloaded, but they are often AOT compiled to a specific ISA, and do not involve JIT compilation at all.)
\end{itemize}

\subsection{Experimental Setup and Steps} \label{sec:experimentsteps}
The experiments were running a Google Pixel 32GB smartphone. It has a Qualcomm Snapdragon 821 64-bit quad-core processor, which implements the ARM big.LITTLE architecture. Two of the four cores have frequency scaling from 0.31 GHz - 1.59 GHz, and the other two scale from 0.31 GHz - 2.15 GHz. We measured the CPU cycles used by each process as the metric to evaluate ShareJIT's and default JIT's performance. Since this metric is sensitive to thread-core configuration and to frequency scaling, we disabled the two ``little'' cores and pinned the two big cores to 1.5 GHz prior to each experiment, in order to make our experimental results repeatable and reproducible.

Each app was run by an automated script, instead of a human operator. The scripts are written with Android ADB shell input commands\footnote{See details about ADB shell input commands at www.raizlabs.com/dev/2017/09/automating-input-events-android/}, which could send events like clicking, swiping, and typing text to the device, simulating a real user's activities on a certain app. Because \textit{Airbnb} detects and prevents robot login, we needed to input a user name and a password manually.
In all other apps, the launch, execution, and data collection were entirely automated. We used automated scripts to eliminate the data deviation that might be caused by human operators, and to maximize the repeatability of the experiments.

We also randomized the order of running apps in each experiment. Because in a sharing relationship, the sharer is always the app used earlier and the sharee is always the app used later, the experimental results for one app depend on the order in which all the apps run. Thus, we randomized the apps running order in each experimental run to amortize the potential result bias that might be caused by any fixed ordering of the apps.

The length of each app's running time ranged from 2 to 5 minutes, depending on the functionalities of the app and what was possible to do from the perspective of a real user. For example, the script for \textit{Chrome} launches the app, inputs a keyword at Google search bar, then browses the items and images for 2 minutes; the script for \textit{Facebook} launches the app, inputs a user name and a password to login, browses the news feed, and browses the market items around the user's location for a total of 5 minutes. 
The amount of work in the script for the same app is consistent across multiple runs.

Each experimental run was performed in the following steps:
\begin{enumerate}\setlength{\itemsep}{0pt} \setlength{\parskip}{3pt}
\item flashing the experimental device, a Google Pixel Phone, with either the default Android OS Image, or the ShareJIT Android OS Image; 
\item installing all 8 benchmark apps;
\item running the 8 automated scripts in a random order;
\item dumping and collecting the data.
\end{enumerate}

The data produced by the experimental runs was, on a per-process basis: compilation time for all the JIT compiled methods, CPU cycles used by the app processes, JIT code cache size and data cache size which includes runtime methods profile information, and stack maps as we mentioned. The JIT compilation time, compiled code size and data size of a method are logged at runtime through the Android Logcat tool\cite{logcat}; the CPU cycles are read from the \textit{/proc/pid/stat} file, which provides status information about the process identified by \textit{pid}. From the \textit{/proc/pid/stat} data, we extracted both user mode and kernel mode cycles consumed by each process. We believe this number, representing the time resources spent by a process, is an accurate and precise metric to compare the performance between ShareJIT and default JIT .

\subsection{Data Noise}
During the experiments, there existed several inevitable sources of data noise. One is some content provider apps served different content from one run to another. For example, \textit{Facebook} and \textit{Instagram} had different posts in each run. And it was impossible to control the content being served in any app or any run. Another source is the inconsistency of the quality/speed of the WiFi network and apps' servers. This effect stands out when there are automatically played videos in an app's content. Higher WiFi network speed or server responding speed brings shorter loading delay of a video before its automatic playing, which leads to more data transfer and communication between an app and its server. To eliminate these noises, we conducted 30 experiments (15 pairs) and took averages; we ran default JIT and ShareJIT alternately and consecutively to minimize the effect of environmental and app-content changes. 

\subsection{Sharing Threshold}\label{sec:sharingthreshold}
As we mentioned in Section \ref{sec:costmodelofsharing}, the benefit of sharing the compiled code for a method is unpredictable because the number of future invocations of a given method is unknowable. So we performed an empirical study to discover the threshold where, in practice, sharing compiled code gains the most performance improvement. 

\begin{figure}[t!]
	%\vspace*{-.5cm}
	\includegraphics[keepaspectratio=true,scale=0.45]{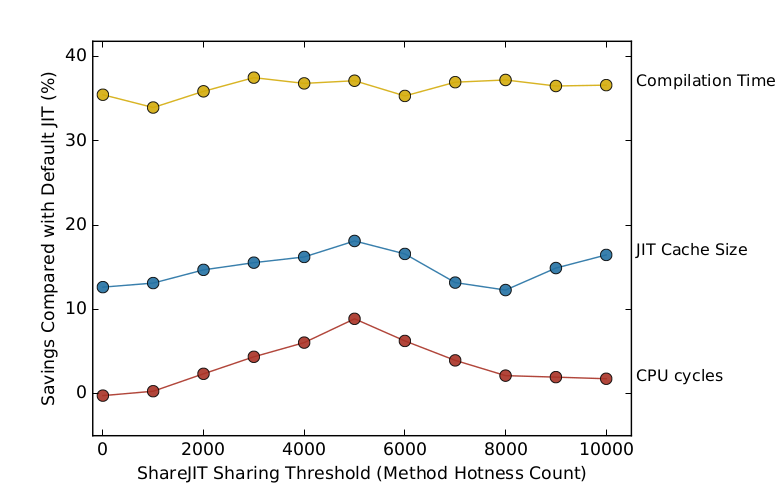}
 	\caption{Savings in CPU Cycles, Compilation Time and JIT Cache Size (including both code cache and data cache), compared with default JIT in ART, as a function of different sharing threshold values}
\label{fig:threshold}
\end{figure}

As mentioned in Section \ref{sec:artjitcompiler}, the threshold for JIT compilation in ART is a hotness count of $10,000$ for a method. At a count of $10,000$, its own JIT compiler will compile it. Thus, the sharing threshold must be lower than $10,000$. Therefore, we divided the range from $1$ to $10,000$ equally and tested the thresholds at 1, 1,000, 2,000, 3,000, \ldots{} , 10,000.
For each tested threshold, we ran the experiment following the procedure described in Section~\ref{sec:experimentsteps}. 
Figure~\ref{fig:threshold} shows the results. The x axis shows different thresholds at which a sharing task is created for a method.
%This sharing task will firstly compute the hash-identification for this method, and then check if the global JIT cache already has corresponding compiled code for it. 
The y axis shows the percentage of savings on compilation time, JIT cache size and CPU cycles, compared with default ART. The values of the data points in this figure are the average results of all 8 apps. 

From this figure, we can see that the compilation time is not significantly statistically correlated with the sharing threshold; all of the compilation time savings fluctuate slightly between 35\% to 37\%.
%It is reasonable since compilation time is only a small part of the time resources consumed by an app.
% NO: the number of methods that get compiled does not change much with the threshold; it only affects
%  the timing of those compilations.
Both cache size and CPU cycles show a peak at a threshold of $5,000$. 
It appears that, for CPU cycles, as the threshold $ST$ increases and the number of methods which qualify for $HC(m) \geq ST$ decreases, both the cost $\sum C(m)$ and the benefit $\sum G(m)$ in equation \ref{eq:total_benefit} decrease; but on the left of the peak, the cost decreases faster, while on the right, the benefit decreases faster. 

For the JIT cache size, it is tempting to think the increased sharing leads to increased code space savings. In truth, however, final code space saving depends on the total number of methods compiled and shared. The value of $ST$ only affects cache size if it changes the set of methods that are ultimately compiled. If a method's hotness count stays in between the sharing threshold and the hot threshold, it will not get compiled when run with either the default system or ShareJIT.
If it does not get compiled in either, then ShareJIT shows no space savings. 
We cannot resolve this problem satisfactorily using arithmetic because the app's behavior is unpredictable. Nevertheless, the empirical study gives us an answer---the best sharing threshold appears to be $5,000$. And the experimental results depicted in the next section are generated with the sharing threshold set to $5,000$.

\subsection{Results and Discussion}\label{sec:results}
The results we show in this section are the average values of all 15 pairs of experiments. If an app consists of multiple processes, e.g., \textit{Chrome} created as many as 6 processes including the main process \texttt{com.chrome}, the sandbox processes such as \texttt{com.chrome:sandboxed\_process0} and the privilege processes such as \texttt{com.chrome:privileged\_process0}, etc., the results include the data from all its processes. 

\def\NT#1{\makebox[\widthof{88.8}][r]{#1}} % right justify in centered field of 88.8
\begin{table}[!b]
\centering\renewcommand{\arraystretch}{1.25}\tabcolsep=7pt
\caption{Code space saving and performance improvements due to ShareJIT compared with default JIT. The bottom row shows average total reductions across all 8 apps.}
\label{tbl:results}

\begin{tabular}{l|ccc|c|c}
    & \multicolumn{3}{c|}{\textbf{JIT Cache Reduction (\%)}} & \textbf{CPU Cycles} 
    & \textbf{Compilation Time}\\[2pt]
\bf{}App &\bf{}Code &\bf{}Data &\bf{}Total &\bf{}Reduction (\%) &\bf{}Reduction (\%)           \\\hline
\it{}Airbnb     & \NT{10.3} & \NT{17.7} & \NT{14.2} & \NT{1.5}  
                & \NT{28.5}\rule{0pt}{14pt} \\
\it{}Amazon     & \NT{9.9}  & \NT{20.1} & \NT{15.2} & \NT{21.9} & \NT{19.1} \\
\it{}Chrome     & \NT{8.0}  & \NT{11.8} & \NT{10.0} & \NT{3.2}  & \NT{21.4} \\
\it{}Facebook   & \NT{22.7} & \NT{26.9} & \NT{24.9} & \NT{13.9} & \NT{57.3} \\
\it{}Firefox    & \NT{3.3}  & \NT{8.4}  & \NT{6.0}  & \NT{3.9}  & \NT{20.1} \\
\it{}Googlemaps & \NT{9.8}  & \NT{17.3} & \NT{13.6} & \NT{1.6}  & \NT{30.7} \\
\it{}Instagram  & \NT{9.2}  & \NT{19.7} & \NT{14.8} & \NT{15.5} & \NT{39.8} \\
\it{}Skype      & \NT{11.3} & \NT{18.9} & \NT{15.0} & \NT{1.7}  & \NT{25.7} \\ \hline
\it{}Average    & \NT{12.7} & \NT{19.7} & \NT{16.4} & \NT{9.0}  & \NT{37.0}
\end{tabular}
\end{table}

Table~\ref{tbl:results} lists all the detailed numbers of our experimental results.
ShareJIT reduces the total JIT code cache by an average of 12.7\% and the total JIT data cache by an average of 19.7\%. Together, ShareJIT reduces the total JIT cache by an average of 16.4\%. From another perspective, ShareJIT saves about $1.9 MB$ space across all 8 apps in an average total running time of 32.5 minutes. ShareJIT consumes about 300 KB space for the global sharing map in an entire experimental run.

\begin{figure*}[t!]
        \centering
        \begin{subfigure}{0.5\textwidth}
            \centering
            \includegraphics[keepaspectratio=true,scale=0.36]{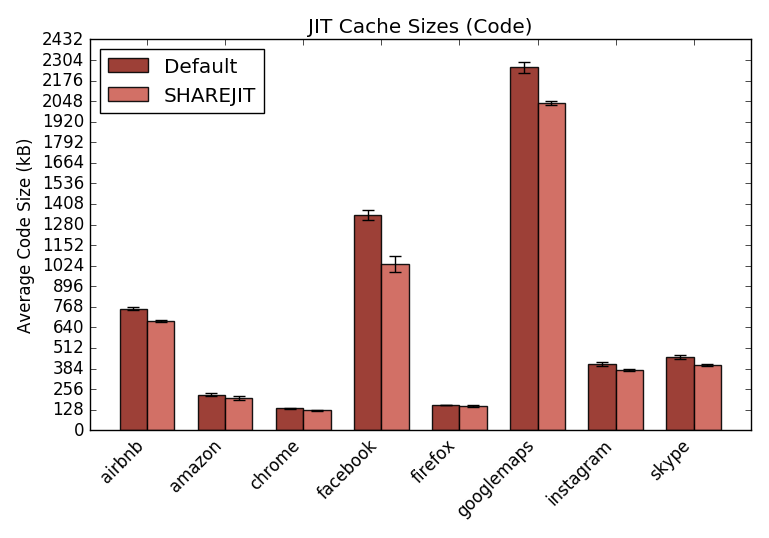}
            %\vspace*{-5pt}
        	%\subcaption{ JIT Code Cache Sizes}
        \end{subfigure}%
        \begin{subfigure}{0.5\textwidth}
            \centering
            \includegraphics[keepaspectratio=true,scale=0.36]{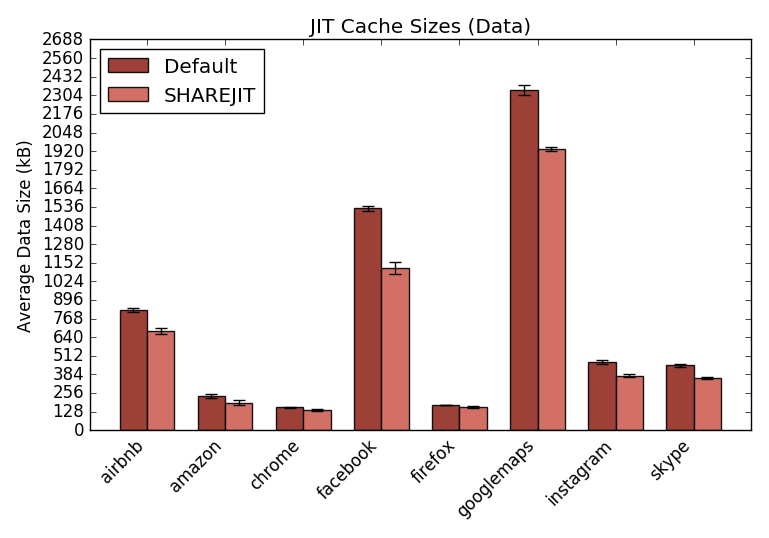}
            %\vspace*{-5pt}
            %\subcaption{JIT Data Cache Sizes}
        \end{subfigure}
        \vspace*{-.3cm}
        \caption{Average end-of-experiment JIT code cache and data cache sizes for each app over all experiments using default JIT and ShareJIT. Error bars show 95\% confidence intervals.}
        \label{fig:jit_size}
    \end{figure*}

Figure~\ref{fig:jit_size} compares the end-of-experiment JIT cache sizes of each app between default JIT and ShareJIT. %The left panel shows the code cache size in each app and the right panel shows the data cache sizes.
More specifically, the left panel of Figure~\ref{fig:jit_size} shows that for every app, ShareJIT decreases the JIT code cache size. The maximum average decrease is in \textit{Facebook}, where ShareJIT reduces the code cache size by an average of 22.7\% (303.9 KB). The right panel of Figure~\ref{fig:jit_size} demonstrates that for JIT data cache, every app also shows reduction by running ShareJIT. The maximum average decrease is still in \textit{Facebook}---26.9\% (410.5 KB). The results of other apps are listed in Table~\ref{tbl:results}.

\begin{figure*}[t!]
        \centering
        \begin{subfigure}{0.5\textwidth}
            \centering
            \includegraphics[keepaspectratio=true,scale=0.36]{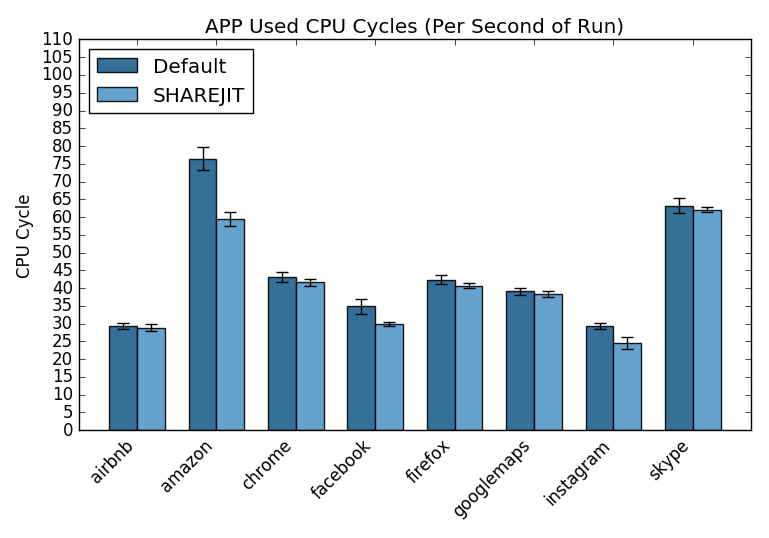}
        	%\subcaption{}
        \end{subfigure}%
        \begin{subfigure}{0.5\textwidth}
            \centering
            \includegraphics[keepaspectratio=true,scale=0.36]{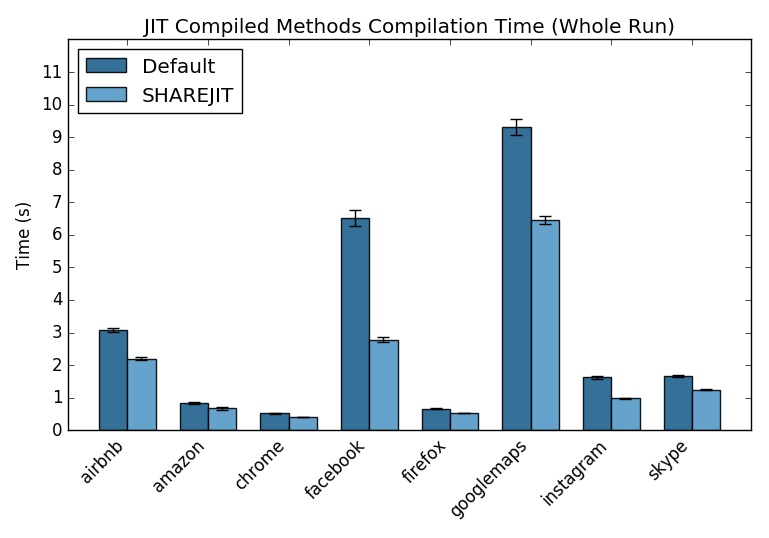}
             %\subcaption{}
        \end{subfigure}
         %\vspace*{-.3cm}
        \caption{Average app used CPU cycles per second of run and average total compilation time of JIT compiled methods for each app over all experiments using default JIT and ShareJIT. Error bars show 95\% confidence intervals.}
        \label{fig:performance}
    \end{figure*}

As we mentioned in Section~\ref{sec:experimentsteps}, we measured the CPU cycles used by each process to represent ShareJIT's and default JIT's performance. The left panel of Figure~\ref{fig:performance} shows the number of CPU cycles used by each app, including both kernel mode and user mode CPU cycles, per second of run. In order to eliminate the effect of different experimental times of different apps, we display the cycles of per second of run instead of the whole run. Thus the computation of the average speedup across all apps is meaningful. From this figure, we could see ShareJIT improves the performance of every app. The maximum speedup percentage appears in \textit{Amazon}---an average of 21.9\%. The average speedup across all apps per second of run is 9.0\%.

The right panel of Figure~\ref{fig:performance} illustrates the total compilation time of JIT compiled methods during run time. It is consumed and logged by the JIT compiler thread in ART. ShareJIT reduces the compilation time of each app. The average total compilation time reduction is 37.0\%. And the maximum total compilation time reduction appears, again, in \textit{Facebook}---an average of 57.3\% (saving 3.7 seconds for a whole run).

    \begin{figure*}[t!]
        \centering
        \begin{subfigure}{0.5\textwidth}
            \centering
            \includegraphics[keepaspectratio=true,scale=0.365]{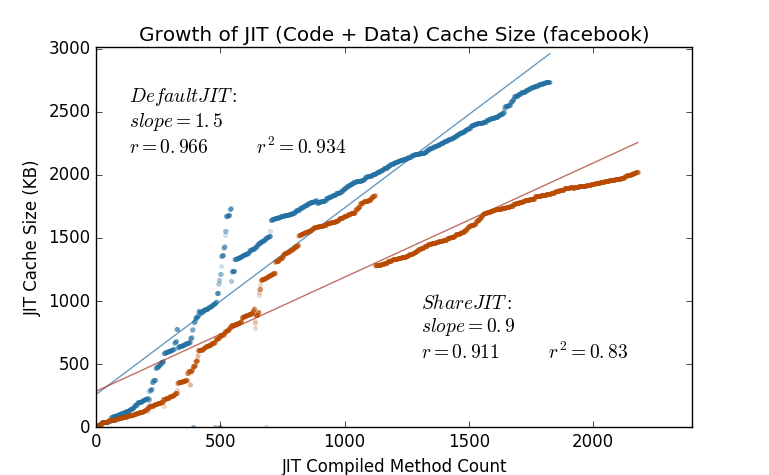}
        	\subcaption{}
        \end{subfigure}%
        \begin{subfigure}{0.5\textwidth}
            \centering
            \includegraphics[keepaspectratio=true,scale=0.365]{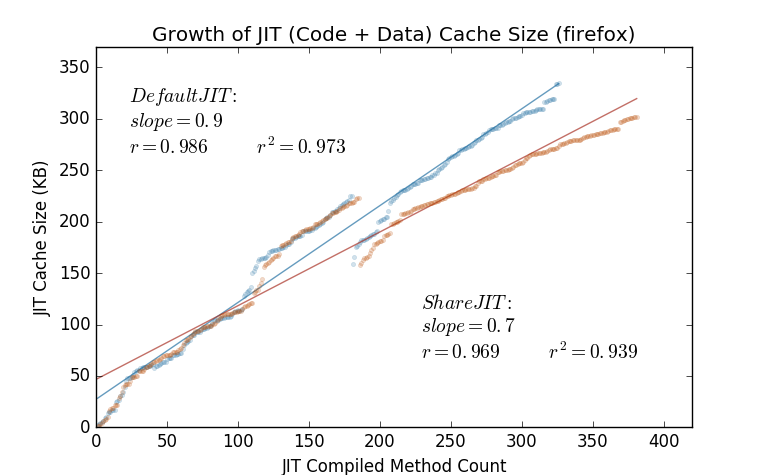}
             \subcaption{}
        \end{subfigure}
  
        \begin{subfigure}{0.5\textwidth}
            \centering
            \includegraphics[keepaspectratio=true,scale=0.365]{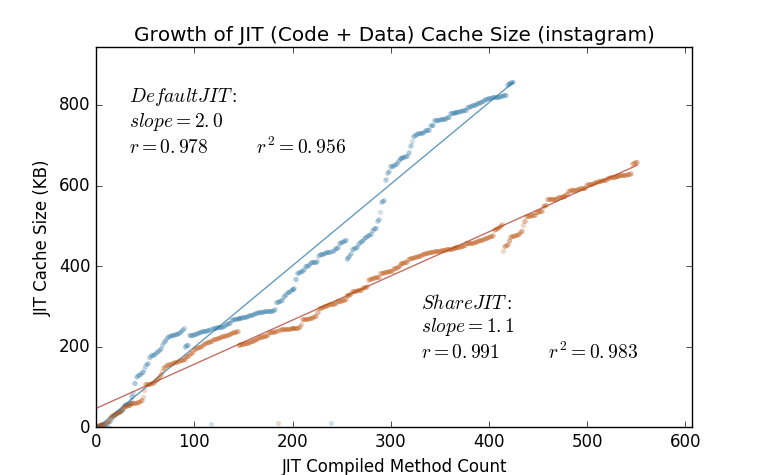}
        	\subcaption{}
        \end{subfigure}%
        \begin{subfigure}{0.5\textwidth}
            \centering
            \includegraphics[keepaspectratio=true,scale=0.365]{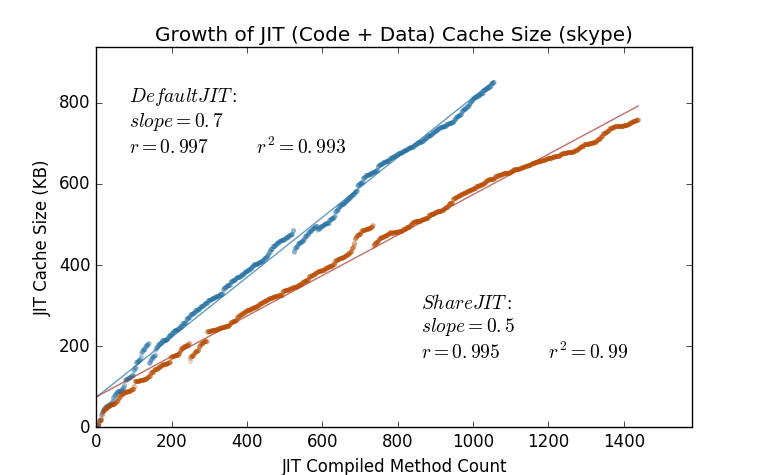}
             \subcaption{}
        \end{subfigure}
         %\vspace*{-.2cm}
        \caption{A scatter plot to show the growth curve of JIT cache size of example apps during a whole experiment run. X axis is the count of JIT compiled methods, and y axis is the corresponding JIT data and cache size. Figure (a)(b)(c)(d) respectively show the growth curve of Facebook, Firefox, Instagram and Skype.}
        \label{fig:growth}
    \end{figure*}
    
Recall that in Section~\ref{sec:memory_utilization}, we've discussed that theoretically, the system-wide memory utilization of ShareJIT is in proportion to the amount of sharing and the size of those methods. However, in practical experiments, the end-of-experiment JIT cache size of each app we measured was actually resulted from a combination of several factors. (`-'' indicates this factor reduces the end-of-experiment cache sizes while ``+'' factors increase them.)

\begin{itemize}
\item[-] ShareJIT eliminates the duplicate copies of compiled code for the same methods existing across apps;
\item[-] Restricted optimizations, e.g. less inline substitution, could reduce the size of a method's compiled code;
\item[+] A ShareJIT garbage collector only collects the compiled code that is not currently being used by any process, so it may delay the collection of shared compiled code, leading to a temporary cache size increase;
\item[+] Given a fixed time frame, ShareJIT could complete more compilation tasks in the JIT thread task queue because each task costs less time compared with default JIT. And more compilations can clearly cause an increase in the end-of-experiment cache size.
\end{itemize}

Figure~\ref{fig:growth} shows the example growth curves of JIT cache sizes during a whole experiment run for 4 apps. In sub-figure (a)(c)(d) which demonstrates the cache size growth of \textit{Facebook}, \textit{Instagram} and \textit{Skype}, the ShareJIT curve (the red one) is consistently below the default JIT curve (the blue one). It is resulted from a combination of both restricted optimization and elimination of duplicate compiled code. All of the four figures also show that ShareJIT compiled more methods than the default JIT did in a given time frame.  For example, in \textit{Skype}, default JIT compiled about 1,100 methods while ShareJIT compiled about 1,500 methods (in a 235 seconds of running). Overall, ShareJIT manages to reduce the JIT cache size by an average of 16.4\%.
    
\section{Related Work}\label{sec:related_work}
The earliest related work we found with regard to code sharing at runtime is \citet{dillenberger2000building}---an implementation of the JVM for OS/390. In this work, multiple JVMs could share class data, constant pools, etc, which are stored in a shared heap during class loading, linking and verification, while the compiled code is not shared. The authors briefly discussed these ideas at a high level without evaluating the actual performance of the system.

\citet{czajkowski2002code} described two systems---one allows the sharing of class meta-data, including bytecodes, among virtual machines, and the other additionally allows the sharing of dynamically compiled code. Both systems were implemented as modifications to the HotSpot JVM client compiler~\cite{kotzmann2008design}. The first one is similar to~\citet{dillenberger2000building}. It divides the heap area into a shared part, a private part and an extra indirection part. Objects in the shared area have to reference objects in the private area via an entry in the indirection table. Each indirection holds the virtual address of the object associated with it, which can be different for each process. The second system uses these mechanisms to share compiled code with little modifications in JIT compilation, e.g. static variables access. Czajkowski et al. pointed out that their systems are less robust/secure because of the use of shared heap.

After Czajkowski et al.'s work, a variety of similar projects have addressed sharing class meta-data, class loaders, and other runtime representations in the JVM~\cite{wong2003dynamically,daynes2005sharing,kuck2009sharing,landau2011shared,berry2004class,schmidt2008sharing,kawachiya2007cloneable,back2000processes,wintergerst2008centralized,bhattacharya2017improving}. Some of them implemented multi-tasking VMs to share runtime memories~\cite{czajkowski2003multi,czajkowski2012multitasking,yan2016using}. In addition, Oracle~\cite{oracle} and IBM~\cite{ibm} also have production Java execution environments, which enable the sharing of class data, e.g., Java class bytecode and class file metadata, across JVMs.
The sharing of class data requires significant modification to the virtual-machine-layer memory-layout, which couples the implementation tightly to a specific runtime system. In particular, for runtime systems like ART, which have a different class file format than the Java standard (one \dex file contains almost all the classes of an app, and symbolic references are globally unique and mutable during runtime), sharing class data would add too much overhead and, thus, be impractical. 

Studies of code sharing across processes have been expanded to persistent code caches~\cite{bruening2008process,guckert2013case}---an effective way to reduce the overhead of dynamic binary translation, which translates binary code from one instruction set architecture to another at runtime, and is often used in system virtualization, system debugging, system security and whole program analysis. These projects have an orthogonal focus to our project's focus.

\citet{huang2010file} did work that is similar to our work. They proposed a native-code sharing mechanism across Dalvik virtual machines \citep{ehringer2010dalvik} on Android 2.1, by storing all the compiled code in a shared file and implementing a daemon process to control the sharing. 
By contrast, ShareJIT composes a global shared cache and allows each process to manage its own cache segment instead of using a centralized agent. 
Since Android 5, the Dalvik VM was replaced by ART, and, in Android 2.1, the dex code of apps was only interpreted before being JIT compiled.
By contrast, Android 7 uses a combination of interpretation, JIT compilation and AOT compilation to execute dex code. 

Intuitively, Android 2.1 could offer more sharing opportunities than Android 7 does since ART has already AOT compiled some shared libraries in the system after its first boot-up. In Android 2.1, the JIT compiler is more like a typical AOT compiler since it generates position independent code with relocation information, while modern JIT compilers often leverage rich runtime context. The experiment shown in~\citet{huang2010file} is preliminary---they tested 10 simple, small-size benchmarks instead of popular real-life apps with billions of downloads as we use; they measured the system performance through a score given by a single benchmark app called \textit{Caffeinemark}. In contrast, we evaluate the performance of ShareJIT by measuring the CPU cycles used by full apps which include both the time spent in executing app methods and the time spent in the runtime system, e.g. compilation, garbage collection, etc.

\section{Conclusion and Future Work}\label{sec:conclusion}
The ShareJIT system provides a new framework for code caching and for sharing cached code across multiple applications in an Android environment. The goal of ShareJIT is to eliminate repeated JIT compilations of the same code and duplicate copies of the resulting compiled code, which occur in existing systems because applications share library code while the runtime system maintains process-private caches. 

ShareJIT provides a global shared cache created by composing the private cache segments and providing coordinated, controlled access, lookup, and memory management across the full set of caches.
The design recognizes that there is a fundamental tradeoff between shareability and optimization; the more optimized the compiled code is, the less shareable it is. 
ShareJIT increases shareability by restraining the amount of runtime context used during JIT compilation. For example, it limits the scope of inlining to increase sharing.
Through ShareJIT, a single compile step can improve the performance of multiple processes and multiple apps. With ShareJIT, a single app can sometimes begin executing compiled code at an earlier point in its progress than it would be in the standard runtime system. 
ShareJIT improves overall system performance while reducing the total amount of memory devoted to caching compiled code and its associated data structures.

We build a practical implementation of ShareJIT in the Android Runtime system, and provide details of that implementation. We use this implementation to show that ShareJIT improves the overall system performance by an average of 9\%, and decreases the memory utilization by an average of 16\%. Additionally, ShareJIT also decreases the amount of time spent on JIT compilation by an average of 37\%.

ShareJIT opens up opportunities for future work.
\begin{itemize}
\item
There are a wide range of management policies to explore. 
For example, we intend to explore more deeply the relationship between the expected costs and benefits of sharing a method so that we could improve the decision making for when to share a method. 
E.g., an intuitive heuristic is that larger methods produce smaller payoff, because it costs more to compute their hash-identifications and they tend to exhibit less dramatic speedup from compiled execution.
\item 
We also plan to experiment on some inter-process coordinations, e.g. maintaining a global sharing method hotness notion together, changing ownership of a method to a sharee process when the sharer/owner dies. Both create a better global perspective of runtime methods, but since they transfer information across the process boundary in Android, they also bring the overhead of inter-process communication and synchronization.
\item We intend to port the Android implementation of ShareJIT to other open-source systems. In fact, ShareJIT was designed and implemented in an easy-to-port style, e.g., ShareJIT does not change the structure of the process-private code cache; it composes the original code cache segments together to form the global cache and gives each process execute access to the entire cache. This minimized changes to the memory management layer and the in-process JIT.
\item
Other subjects of possible interests include eviction and garbage collection policies, data mining ShareJIT's behavior to identify methods that tend not to be shared and should, thus, be compiled for performance rather than shareability.
\end{itemize}

\bibliographystyle{ACM-Reference-Format}
\balance
\bibliography{references.bbl}

\end{document}